\documentclass[letter]{ptptex}

\usepackage{graphicx}
\usepackage{wrapft}

\def\lesim{\m@thcombine<\sim}
\def\gesim{\m@thcombine>\sim}
\def\lessgtr{\m@thcombine<>}
\def\gtrless{\m@thcombine><}

\newcommand{\Se}{${}^{68}$Se}

\newcommand{\bra}[1]{\left\langle #1 \right|}
\newcommand{\ket}[1]{\left| #1 \right\rangle}
\newcommand{\Hhat}{\hat{H}}
\newcommand{\HhatMq}{\hat{H}_M(q)}

\newcommand{\Hc}{{\cal H}}
\newcommand{\Nhat}{\hat{N}}

\newcommand{\Qhat}{\hat{Q}}

\newcommand{\Phat}{\hat{P}}

\newcommand{\phiqppn}{\phi(q,p,\varphi,N)}
\newcommand{\phiqpn}{\phi(q,p,N)}

\newcommand{\phiq}{\phi(q)}

\newcommand{\del}{\partial}

\newcommand{\beq}{\begin{equation}}
\newcommand{\beqa}{\begin{eqnarray}}
\newcommand{\eeq}{\end{equation}}
\newcommand{\eeqa}{\end{eqnarray}}

\title{
Collective Path Connecting the Oblate and Prolate Local Minima 
in \Se
}

\author{Masato KOBAYASI,$^1$~~ Takashi NAKATSUKASA,$^2$~~ 
Masayuki MATSUO$^3$ \\ 
and~ Kenichi MATSUYANAGI$^1$}

\inst{
{\small\it $^1$ Department of Physics, Graduate School of Science,}\\
{\small\it Kyoto University, Kitashirakawa, Kyoto 606-8502, Japan}\\
{\small\it $^2$ Institute of Physics and Center for Computational Science, 
University of Tsukuba,}\\
{\small\it Tsukuba 305-8571, Japan}\\
{\small\it $^3$ Graduate School of Science and Technology,}\\
{\small\it Niigata University, Niigata 950-2181, Japan}
}

\abst{
By means of the adiabatic self-consistent collective coordinate method
and the pairing plus quadrupole interaction, we have obtained
the self-consistent collective path connecting the oblate and prolate 
local minima in \Se~for the first time. 
Result of calculation indicates importance of 
triaxial deformation dynamics in the oblate-prolate
shape coexistence phenomena.
}

\begin{document}

\maketitle

Shape coexistence phenomena are typical examples of large 
amplitude collective motion in nuclei.   
These phenomena imply 
that different solutions of the Hartree-Fock-Bogoliubov (HFB) equations
(local minima in the deformation energy surface) appear 
in the same energy region and that the nucleus exhibits 
large amplitude collective motion connecting these different
equilibrium points. The identities and mixings of these different shapes
are determined by the dynamics of such collective motion. 
Some years ago, we have proposed a new method of describing 
such large-amplitude collective motion, 
which is called Adiabatic Self-Consistent 
Collective Coordinate (ASCC) method\cite{mat00}.
It yields a new method of solving the basic equations 
of the SCC method\cite{mar80} 
using an expansion in terms of the collective momentum.
It does not assume a single local minimum, so that it is expected to be
suitable for the description of the shape coexistence phenomena. 
The ASCC method also enables us to include the pairing correlations 
self-consistently, removing the spurious number fluctuation modes.
To examine the feasibility of the ASCC method,
we have first applied it to an exactly solvable model called
the multi-$O(4)$ model, which is a simplified version of 
the pairing-plus-quadrupole (P+Q) interaction model\cite{bar68}.
It is shown that the method yields a faithful 
description of tunneling motion through a barrier between
the prolate and oblate local minima 
in the collective potential\cite{kob03}.

In this Letter, we give a brief report of our first application of
the ASCC method to a realistic P+Q interaction model. 
We illustrate its practicality
taking as a typical example the oblate-prolate shape coexistence phenomenon
in $^{68}$Se recently observed in experiments.\cite{fis00} 
The self-consistent collective path obtained successfully 
by means of the ASCC method
is found to run approximately along the valley
connecting the oblate and prolate local minima 
in the collective potential energy landscape. 
To the best of our knowledge,
this is the first time that a self-consistent collective path 
is obtained for realistic situation starting from the
microscopic P+Q Hamiltonian.
We note that a similar approach to large amplitude collective motions
was recently pursued by Almehed and Walet\cite{alm04}, 
although they discussed different nuclei and 
encountered some difficulties in obtaining 
self-consistent collective paths. 

We assume that large-amplitude collective motions are described 
by a set of time-dependent HFB state vectors $\ket\phiqppn$
parametrized by a single collective coordinate $q$,
the collective momentum $p$ conjugate to $q$,
the particle number $N$ and the gauge angle $\varphi$ conjugate to $N$.
As discussed in Ref.~\citen{mat00}, the state vector can be written
\begin{equation} 
\ket{\phiqppn}=e^{-i \varphi \Nhat}\ket{\phiqpn}
=e^{-i \varphi \Nhat}e^{ip{\hat Q(q)}}|\phi(q)\rangle.
\end{equation}
Making an expansion with respect to $p$ and requiring that 
the time-dependent variational principle be fulfilled 
up to the second order in $p$,
we obtain the following set of equations to
determine $\ket{\phiq}$, the infinitesimal generator
$\Qhat(q)$, and its canonical conjugate  $\Phat(q)$:

\begin{equation} 
\delta\bra{\phiq}\HhatMq \ket{\phiq} = 0,   
\label{eqcsmf}
\end{equation}

\begin{equation} 
\delta\bra{\phiq}[\HhatMq, \Qhat(q) ] - {1\over i} B(q) \Phat(q)
\ket{\phiq} = 0,   
\label{eqcshq}
\end{equation}

\begin{equation}
\delta\bra{\phiq} [\HhatMq, {1\over i}\Phat(q)] -C(q)\Qhat(q) 
-{1 \over 2B(q)}[[\HhatMq, (\Hhat - \lambda(q)\Nhat)_{A}], \Qhat(q)]
-{\del \lambda \over \del q}\Nhat
\ket{\phiq} = 0.  
\label{eqcshp}
\end{equation}

\noindent
Here 
\begin{equation}
\HhatMq=\hat{H}-\lambda(q)\hat{N}-\frac{\partial V}{\partial q}\hat{Q}(q)
\end{equation}
is the Hamiltonian in the moving frame; 
\begin{equation}
C(q) = {\del^2 V \over \del q^2} 
+ {1\over 2B(q)}{\del B\over \del q}{\del V \over \del q}
\end{equation}
is the local stiffness;
\noindent
$(\Hhat - \lambda\Nhat)_{A}$ 
represents the two-quasiparticle creation and
annihilation parts of ($\Hhat - \lambda\Nhat$);
$\Qhat(q)$ and $\Phat(q)$ satisfy the canonical variable condition
\begin{equation}
\bra{\phiq}[\Qhat(q),\Phat(q)]\ket{\phiq} = i. \label{cvcqp} 
\end{equation}
\noindent
Once $\ket{\phiq}$ and the infinitesimal generators are determined
for every values of $q$,   
we obtain the collective Hamiltonian 
$\Hc(q,p) = {1\over 2} B(q) p^2 + V(q)$
with the collective potential
$V(q) = \bra{\phiq}\Hhat\ket{\phiq}$  
and the inverse mass 
$B(q) = -\bra{\phiq}[[\Hhat,\Qhat(q)],\Qhat(q)]\ket{\phiq}$.

%%%%%%%%%%%%%%%%%%%%%%%%%%%%%%%%%%%%%%%%%%%%%%%%%%%%%%%%%%%%%%%%%%%%%%%%%%
\vspace{5mm}
\begin{center}
{\footnotesize Table I.~~  
Spherical single-particle orbits and their energies used in the calculation.\\
Energies relative to those of $1g_{9/2}$ are written in MeV.
}    
\begin{tabular}{cccccccccc} \hline \hline
orbits 
& $1f_{7/2}$ &$2p_{3/2}$ & $1f_{5/2}$   
& $2p_{1/2}$ &$1g_{9/2}$ & $2d_{5/2}$ 
& $1g_{7/2}$ &$3s_{1/2}$ & $2d_{3/2}$ \\ \hline
protons    
&  -8.77     &  -4.23    & -2.41  
&  -1.50     &   0.0     &  6.55  
&   5.90     &  10.10    &  9.83  \\
neutrons   
&  -9.02     &  -4.93    & -2.66  
&  -2.21     &   0.0     &  5.27  
&   6.36     &   8.34    &  8.80  \\ \hline
\end{tabular}
\end{center}
\vspace{10mm}

%%%%%%%%%%%%%%%%%%%%%%%%%%%%%%%%%%%%%%%%%%%%%%%%%%%%%%%%%%%%%%%%%%%%%%%%%%

We use the P+Q interaction model 
with the prescriptions of Ref.~\citen{bar68}
for the microscopic Hamiltonian $\Hhat$, 
but here the pairing and quadrupole force parameters are chosen as
$G= 0.320$ MeV (for both protons and neutrons) and $\chi'= 0.248$ MeV
so that the constrained HFB potential energy surface
(shown by contour lines in Fig.1) exhibits two local minima at
prolate and oblate shapes, whose pairing gaps, quadrupole
deformation and energy difference approximately reproduce 
those obtained in a recent Skyrme-HFB calculation 
by Yamagami {\it et al.}\cite{yam01}.
The spherical single-particle energies are taken from
those of the modified oscillator model of Ref.~\citen{ben85} 
and listed in Table I.
In this way the effective Hamiltonian provides a suitable situation 
with which shape coexistence dynmamics can be studied,
although further improvements, {\it e.g.}, by including the quadrupole pairing
and/or neutron-proton pairing interactions, may better be taken into
account for quantitative comparison with experimental data.

%%%%%%%%%%%%%%%%%%%%%%%%%%%%%%%%%%%%%%%%%%%%%%%%%%%%%%%%%%%%%%%%%%%%%%%%

We have used the following algorithm to solve the
set of ASCC equations (\ref{eqcsmf}), (\ref{eqcshq}), (\ref{eqcshp}) and
(\ref{cvcqp}).
Let the state vector $\ket{\phiq}$ be known at a specific value of $q$. 
We first solve the local harmonic equations in the moving frame
(the moving frame RPA), (\ref{eqcshq}) and (\ref{eqcshp}), 
under the condition (\ref{cvcqp}) 
to obtain $\Qhat(q)$ and $\Phat(q)$.
This is done by a straightforward extension of the procedure 
described in Ref.~\citen{kob03}.
We then construct a state vector at the neighboring point $q+\delta q$
by using the infinitesimal generator $\Phat(q)$ as
\begin{equation}
|\phi(q+\delta q)\rangle = e^{-i \delta q {\hat P(q)}}|\phi(q)\rangle,
\label{deltaq}
\end{equation}
and solve the moving frame RPA with respect to this state 
to obtain $\Qhat(q+\delta q)$ and $\Phat(q+\delta q)$.
Though the above $|\phi(q+\delta q)\rangle$ does not necessarily satisfy
the HFB equation in the moving frame (\ref{eqcsmf}),
%because we use a finite step $\delta q$ although the
%infinitesimal generator $\hat P(q)$ continuously changes as $q$ varies.  
we can use this state vector as an initial solution of (\ref{eqcsmf}) 
at $q+\delta q$. 
We search for the solution of (\ref{eqcsmf}) under the constraints
\begin{eqnarray}
&& \langle\phi(q+\delta q)|\Nhat|\phi(q+\delta q)\rangle = N, \label{constr1} \\
&& \langle\phi(q+\delta q)|\Qhat(q)|\phi(q+\delta q)\rangle = \delta q
\label{constr2}
\end{eqnarray}
by means of the gradient method.
Here the nucleon-number constraint (\ref{constr1}) is actually applied
for both proton and neutron numbers. 
Equation (\ref{constr2}) is the constraint for 
the increment $\delta q$ of the collective coordinate.
After finding a solution of Eq. (\ref{eqcsmf}), we renew
$\hat Q(q+\delta q)$ and $\hat P(q+\delta q)$ 
by solving again the moving frame RPA equations,
(\ref{eqcshq}) and (\ref{eqcshp}), for the
new state vector $|\phi(q+\delta q)\rangle$ obtained above.
Then we again solve Eq. (\ref{eqcsmf}) with the renewed $\hat Q(q+\delta q)$.
If the above iterative procedure converges, we get the selfconsistent 
solutions that satisfy Eqs.  (\ref{eqcsmf}), (\ref{eqcshq}), (\ref{eqcshp}) and
(\ref{cvcqp}) simultaneously at $q+\delta q$, and we can proceed to 
the next point $q+2\delta q$.
In actual numerical calculation, we start the procedure 
from one of the HFB local minimum and examine whether we arrive 
at the other local minimum by going along
the collective path obtained above. 
We have checked that the same collective path is
obtained by starting from the other local minimum and 
proceeding in an inverse way.

%%%%%%%%%%%%%%%%%%%%%%%%%%%%%%%%%%%%%%%%%%%%%%%%%%%%%%%%%%%%%%%%%%%%%%%%%%%%%%
\begin{figure}[htb]
\parbox{\halftext}{%   %\def\halftext{.471\textwidth}
%\figurebox{6cm}{2cm}
\centerline{\includegraphics[width=6 cm, height=4.5 cm]{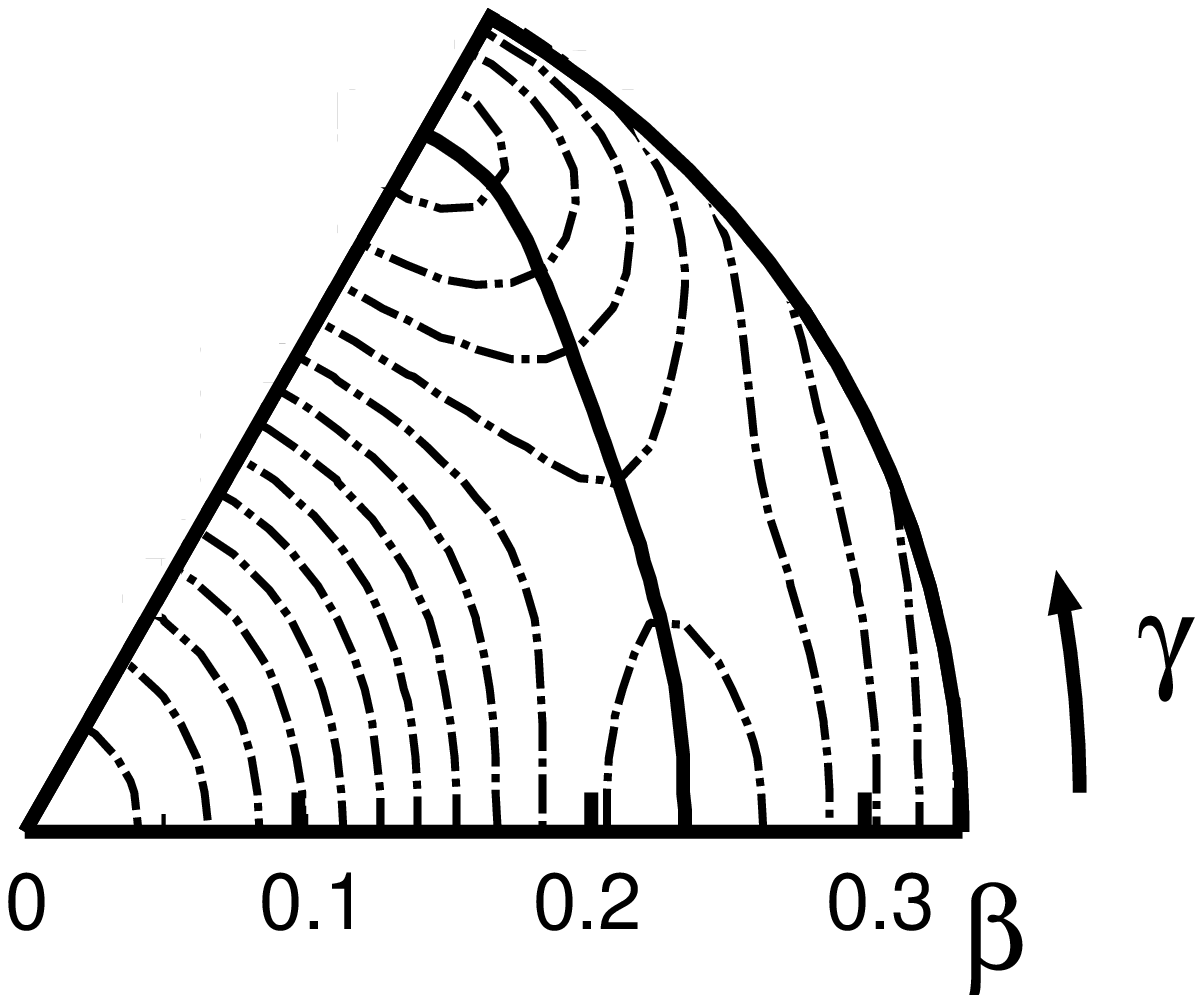}}
\caption{
The bold curve represents the ASCC path connecting the oblate 
and prolate minima in $^{68}$Se projected on the $(\beta,\gamma)$ plane. 
The contour lines are calculated by the conventional
constrained HFB method and plotted for every 100 keV. 
}}
\hfill
\parbox{\halftext}{
\centerline{\includegraphics[width=6 cm, height=4.5 cm]{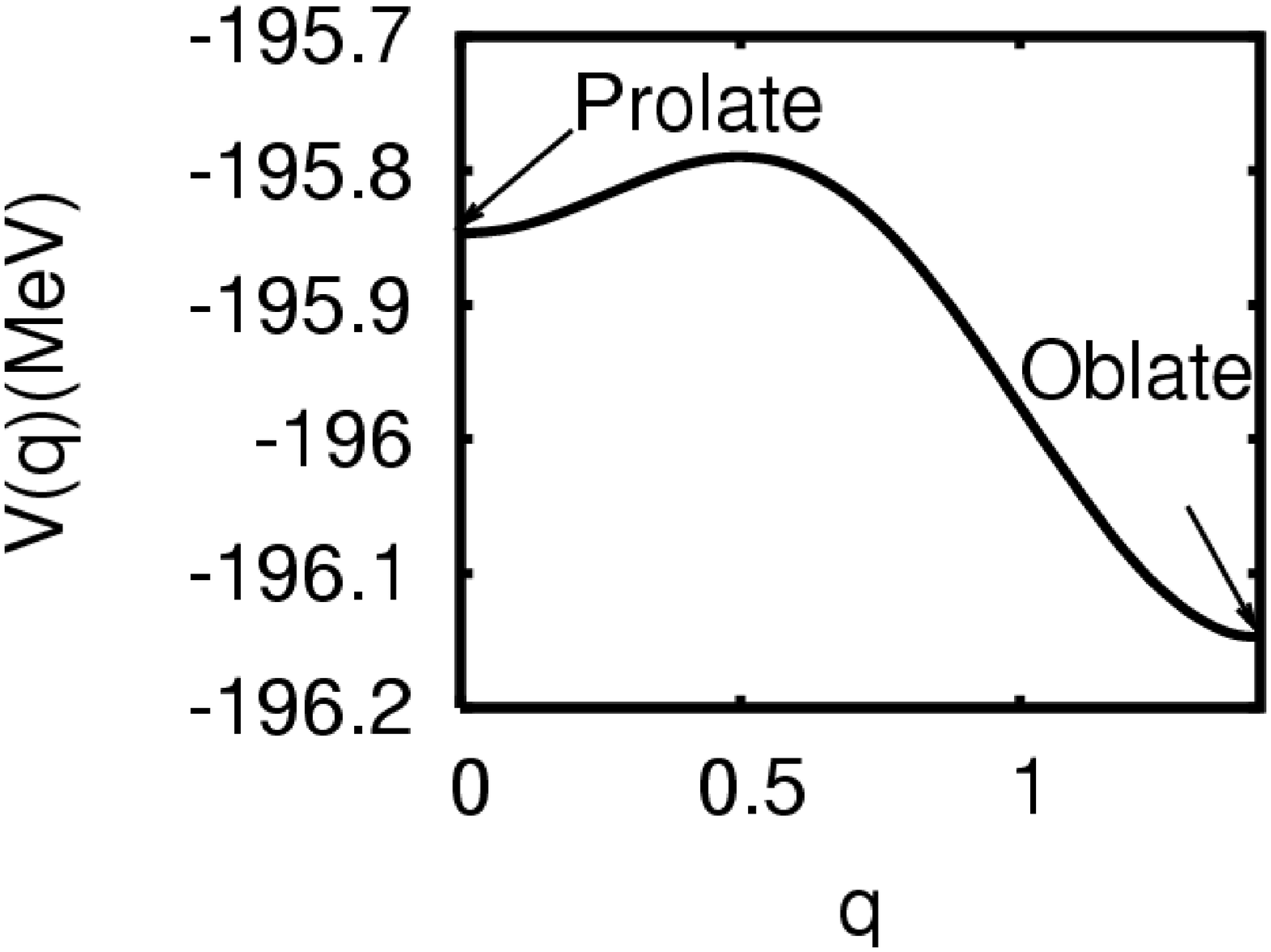}}
\caption{
Collective potential $V(q)$
plotted as a function of the collective coordinate $q$.
Here the origin of $q$ is chosen to coincide with the prolate local minimum 
and its scale is defined such that the collective mass $M(q)=1$~MeV$^{-1}$.
%Due to the symmetry 
%$V(\beta,\gamma)=V(\beta,-\gamma)=V(\beta,\gamma+\frac{2\pi}{3})$, 
%$V(q)$ is symmetric with respect to the oblate minimum $(q=4.5)$ 
%and the two prolate minima (at $q$=0 and 9) represent the same shape.
}}
\end{figure}
%%%%%%%%%%%%%%%%%%%%%%%%%%%%%%%%%%%%%%%%%%%%%%%%%%%%%%%%%%%%%%%%%%%%%%%%%%

%%%%%%%%%%%%%%%%%%%%%%%%%%%%%%%%%%%%%%%%%%%%%%%%%%%%%%%%%%%%%%%%%%%%%%%%%%%%%%
\begin{figure}[htb]
\parbox{\halftext}{%   %\def\halftext{.471\textwidth}
%\figurebox{6cm}{2cm}
%\figurebox{6cm}{2cm}
\centerline{\includegraphics[width=6 cm,height=6 cm]{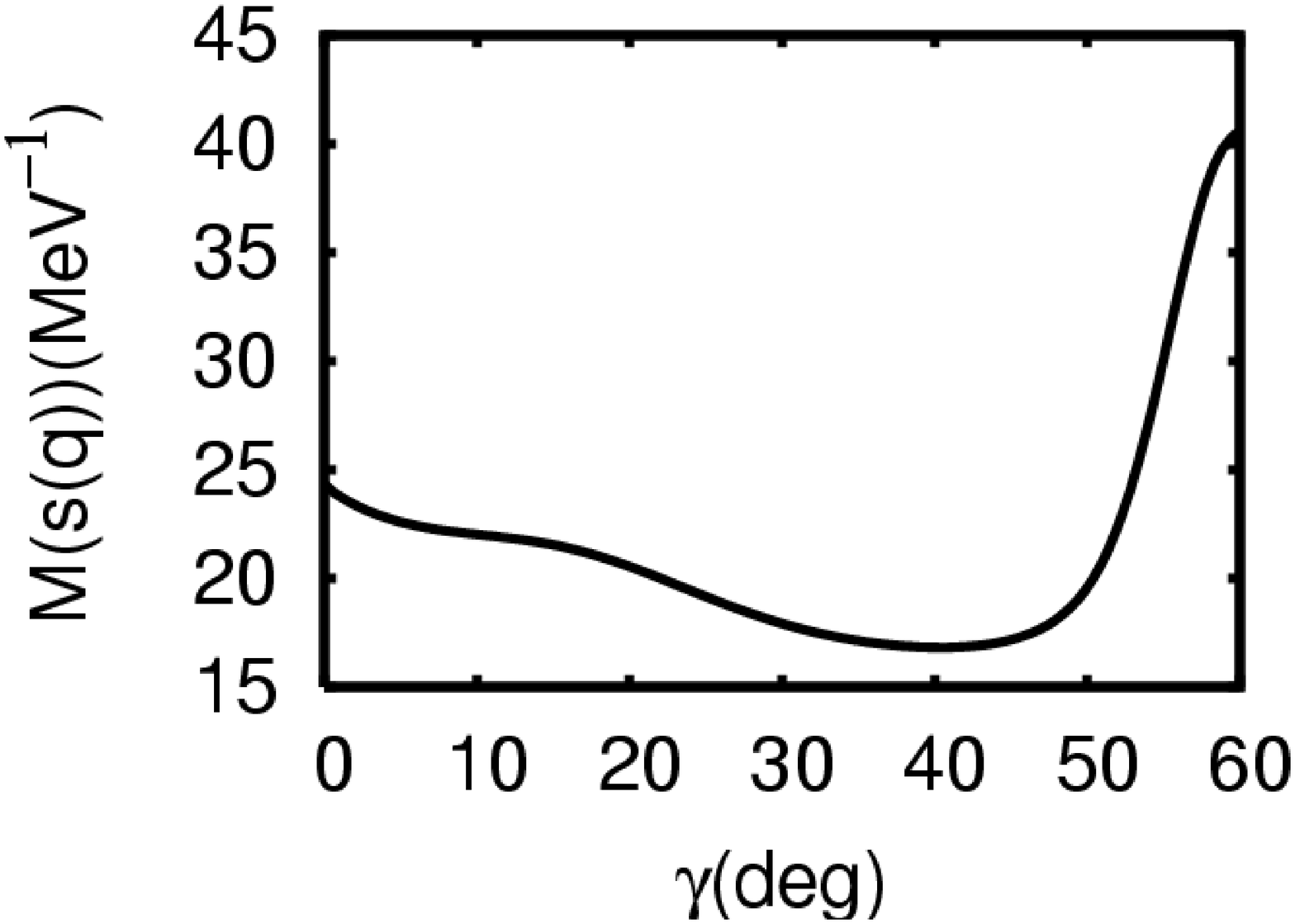}}
\caption{
Collective Mass $M(s)$ with respect to the
geometrical length $s$ along the collective path 
in the $(\beta, \gamma)$ plane is plotted
as a function of the triaxiality parameter $ \gamma$.
}}
\hfill
\parbox{\halftext}{
%\figurebox{6cm}{5cm}
\centerline{\includegraphics[width=6 cm,height=6 cm]{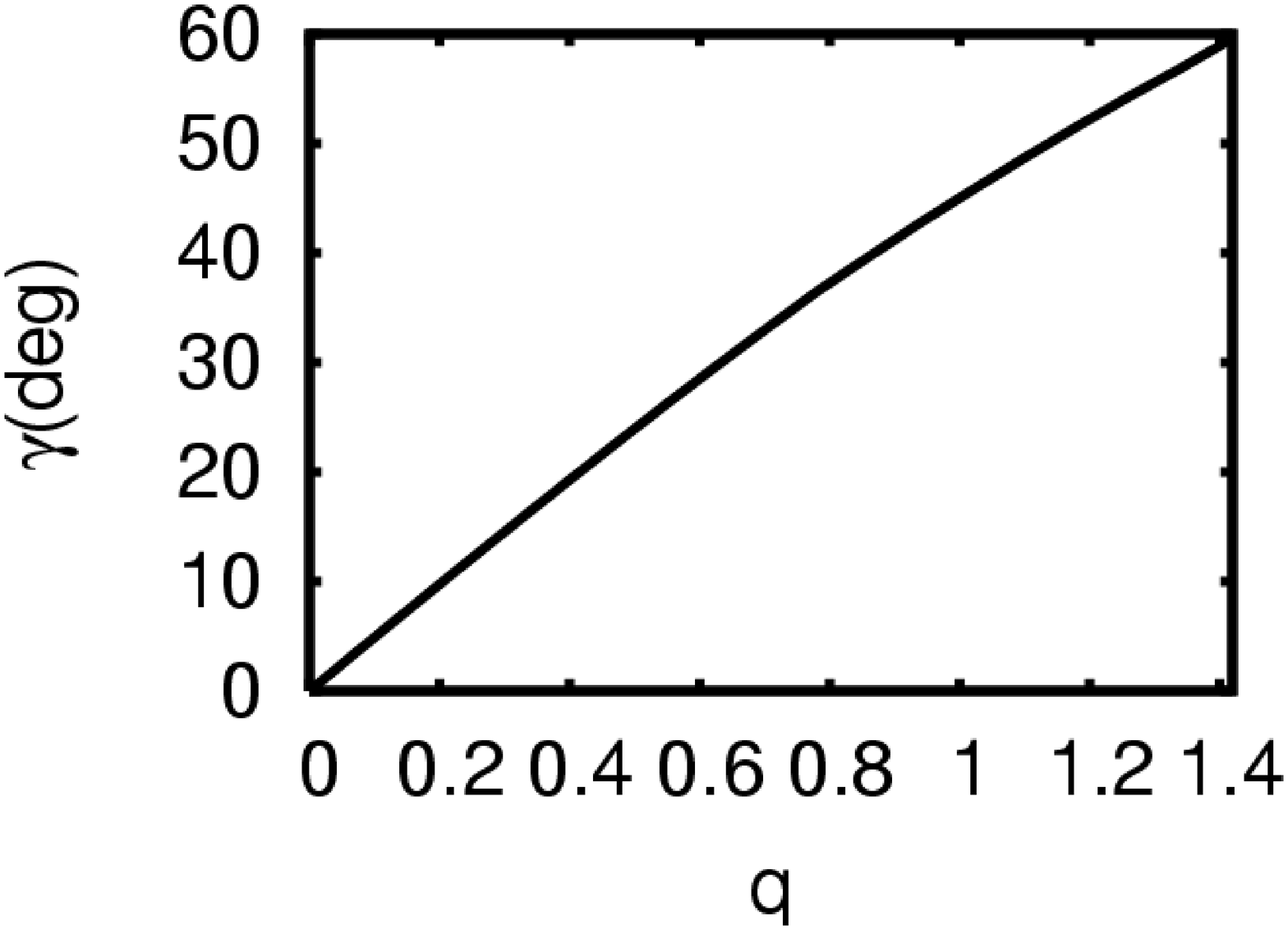}}
\caption{
The triaxiality parameter $\gamma$ 
plotted as a function of the collective coordinate $q$.
}}
\end{figure}
%%%%%%%%%%%%%%%%%%%%%%%%%%%%%%%%%%%%%%%%%%%%%%%%%%%%%%%%%%%%%%%%%%%%%%%%%%

%%%%%%%%%%%%%%%%%%%%%%%%%%%%%%%%%%%%%%%%%%%%%%%%%%%%%%%%%%%%%%%%%%%%%%%%%%%%%%
\begin{figure}[htb]
\parbox{\halftext}{%   %\def\halftext{.471\textwidth}
%\figurebox{6cm}{2cm}
\centerline{\includegraphics[width=6 cm,height=6 cm]{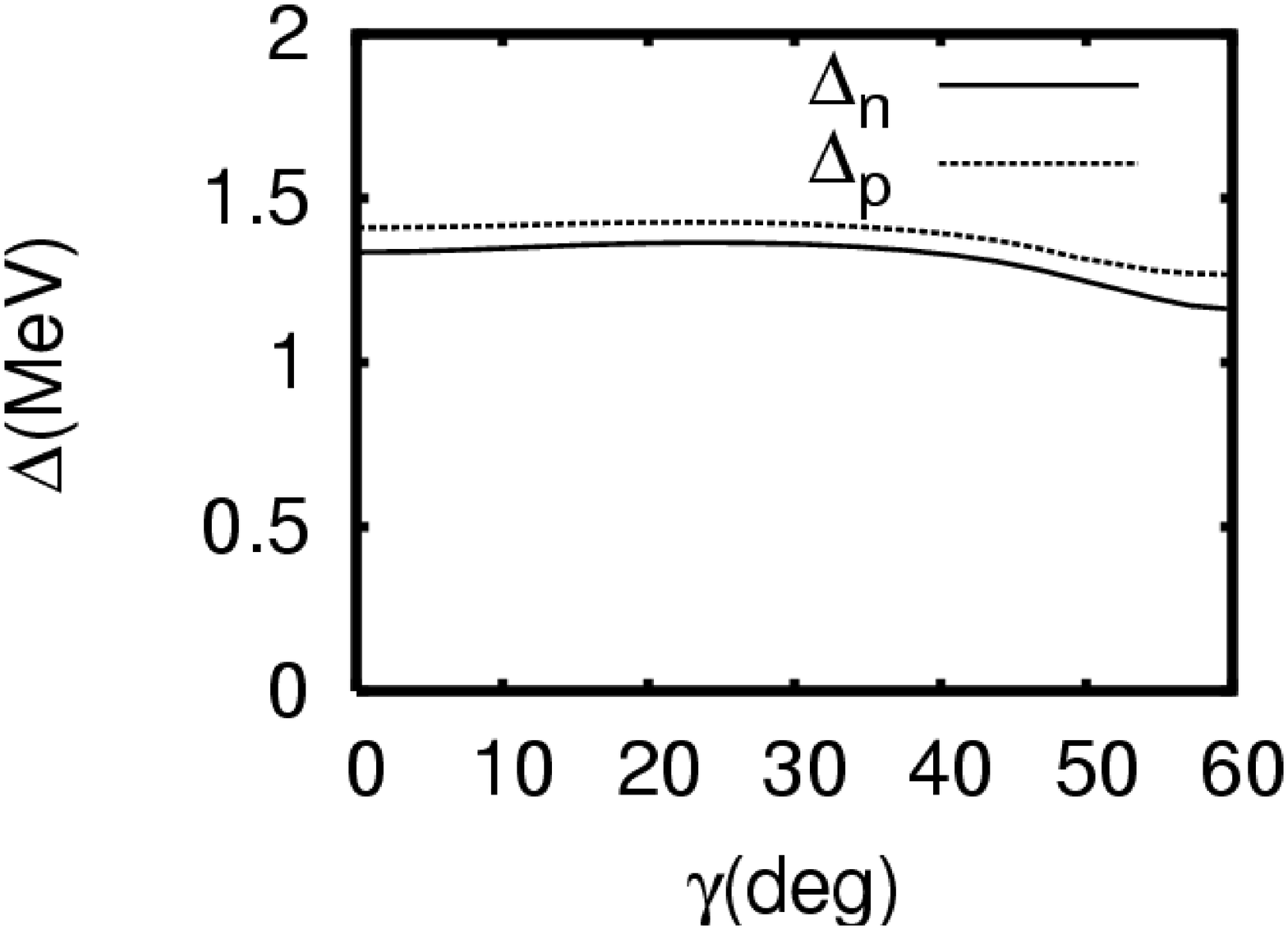}}
\caption{
Neutron and proton pairing gaps, $\Delta_n$ and $\Delta_p$, 
plotted as functions of $\gamma$.
}}
\hfill
\parbox{\halftext}{
%\figurebox{6cm}{5cm}
\centerline{\includegraphics[width=6 cm,height=6 cm]{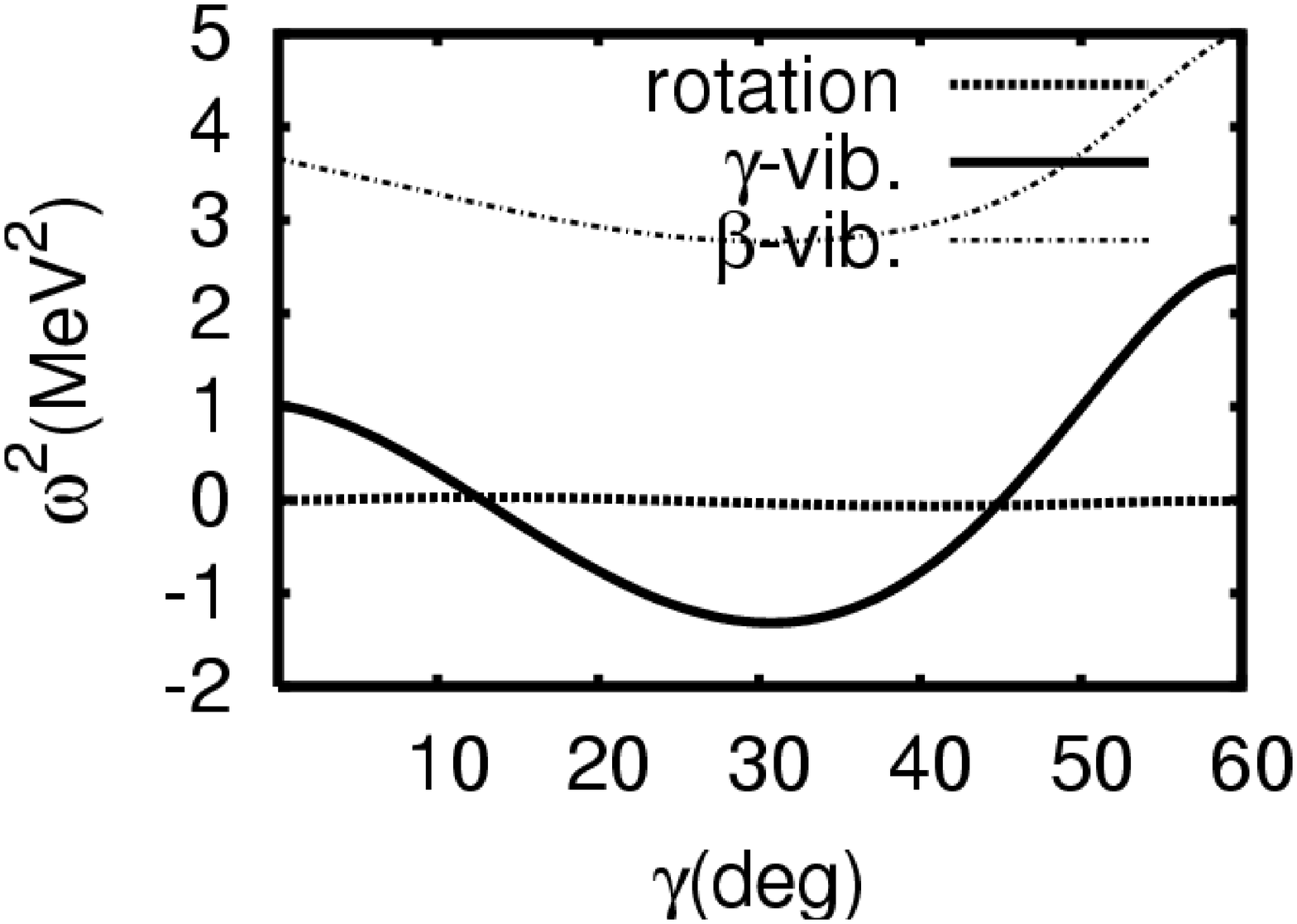}}
\caption{
Lowest three eigen-frequencies squared, $\omega^2=BC$,
of the moving frame RPA, 
plotted as functions of $\gamma$.
}}
\end{figure}
%%%%%%%%%%%%%%%%%%%%%%%%%%%%%%%%%%%%%%%%%%%%%%%%%%%%%%%%%%%%%%%%%%%%%%%%%%

Carrying out the above procedure we have successfully obtained the
collective path connecting the oblate and prolate local minima in \Se.
The result is shown in Fig.~1.
The deformation parameters $\beta$ and $\gamma$ are here defined as usual 
through the expectation values of the quadrupole operators.\cite{yam01}
Roughly speaking, the collective path goes through 
the valley that exists in the $\gamma$ direction
and connects the oblate and prolate minima. 
If one treats the $\beta$ as collective coordinate and
connects the oblate and prolate shapes through the spherical point,
variation of the potential energy would be much greater 
than that along the collective path we obtained.
The potential energy curve $V(q)$ along the collective path
evaluated by the ASCC method is shown in Fig.~2.
We have defined the scale of the collective coordinate $q$ 
such that the collective mass $M(q)=B(q)^{-1}=1$~MeV$^{-1}$. 
The collective mass as a function of the geometrical length $s$ 
along the collective path in the $(\beta,\gamma)$ plane may be defined
by $M(s)=M(q)(ds/dq)^{-2}$ with $ds^2=d\beta^2+\beta^2d\gamma^2$.
This quantity is presented in Fig.~3 as a function of $\gamma$.
The triaxial deformation parameter $\gamma$ is plotted 
as a function of $q$ in Fig.~4.
Variations of the pairing gaps and of the lowest few eigen-frequencies 
of the moving frame RPA along the collective path
are shown in Figs.~5 and 6. 
The solid curve in Fig.~6 represents the squared frequecy
$\omega^{2}(q)=B(q)C(q)$,
given by the product of the inverse mass $B(q)$ and
the local stiffness $C(q)$, for the moving frame RPA mode 
that develops from the $\gamma$-vibration
in the oblate and prolate limits 
and determines the infinitesimal generators $\Qhat(q)$ and $\Phat(q)$
along the collective path.
The other two curves are solutions of the moving frame RPA having
characters of the collective rotational motion and the $\beta$-vibration,
which are however irrelevant to the collective path.
Note that the frequency of the $\gamma$ mode becomes imaginary
in the region $12^{\circ} < \gamma < 45^{\circ}$.
These results will reveal interesting dynamical properties of 
the shape coexistence phenomena in $^{68}$Se. 
For instance, the large collective mass
in the vicinity of $\gamma=60^\circ$ (Fig. 3) might increase stability of
the oblate shape in the ground state. 
Detailed discussions on these quantities  
as well as the solutions of the collective Schr\"odinger equation 
will be given in a forthcoming full-length paper\cite{kob04}.

In summary, we have applied the ASCC method
to the oblate-prolate shape coexistence phenomena in \Se.
It was found that the collective path goes through the 
valley of the potential energy landscape in the $(\beta,\gamma)$ plane 
along which  the triaxial deformation 
parameter $\gamma$ changes between $0^{\circ}$ and $60^{\circ}$ 
keeping the axially symmetric deformation parameter $\beta$ 
approximately constant.
This is the first time that a self-consistent
collective path between the oblate and prolate minima is obtained
for the realistic P+Q interaction model.
Currently, the generater coordinate method has often been used
to describe variety of shape coexistence phenomena taking the
$\beta$ as the generater coordinate\cite{dug03}. 
The triaxial shape vibrational degrees of freedom is ignored
also in the extensive microscopic calculation of Ref.~\citen{pet02}.
The result of the ASCC calculation, however, strongly indicates 
the necessity of taking into account the $\gamma$ degree of 
freedom at least for describing the oblate-prolate shape coexistence
in $^{68}$Se. 
Effects of triaxial deformation dynamics on various properties 
of shape coexistence, including results of calculation for 
neighboring nuclei, will be discussed in a full-length paper
\cite{kob04}.

We thank Drs. D. Almehed and N. R. Walet for useful
discussions and kindly pointing out a graphical error
in the preprint version of this Letter.
The numerical calculations were performed on the NEC SX-5 supercomputer
at Yukawa Institute for Theoretical Physics, Kyoto University.
This work was supported by the Grant-in-Aid  for Scientific
Research (Nos.~13640281, 14540250 and 14740146) 
from the Japan Society for the Promotion of Science.

\end{document}